\documentclass[letterpaper, 10 pt, journal]{IEEEtran}
\usepackage{cite}
\usepackage{amsmath,amssymb,amsfonts}
\usepackage{algorithmic}
\usepackage{graphicx}
\usepackage{textcomp}

\usepackage{cite}
\usepackage{amsmath,amssymb,amsfonts}
\usepackage{algorithmic}
\usepackage{textcomp}
\usepackage{graphicx,color}
\usepackage{mathrsfs}
\usepackage[vlined,ruled]{algorithm2e}
\usepackage{subfigure}
\usepackage{url}
\usepackage{color}
\usepackage{dsfont}
\usepackage{bbm}
\usepackage{booktabs}
\usepackage{array}
\usepackage[table]{xcolor} 
\usepackage{yfonts}
\usepackage[normalem]{ulem}
\usepackage{arydshln}
\usepackage{multicol}
\usepackage{amsmath}
\usepackage{stfloats}

  \newenvironment{smallarray}[1]
 {\null\,\vcenter\bgroup\scriptsize
  \arraycolsep=.13885em
  \hbox\bgroup$\array{@{}#1@{}}}
 {\endarray$\egroup\egroup\,\null}

\newtheorem{theorem}{Theorem}[section]
\newtheorem{lemma}[theorem]{Lemma}
\newtheorem{definition}{Definition}

\newtheorem{problem}{Problem}
\newtheorem{remark}{Remark}
\newtheorem{assumption}[theorem]{Assumption}

\newtheorem{appxlem}{Lemma}[subsection]

\usepackage{tabularx}
\usepackage{multirow}
\usepackage{makecell}

\newcommand\aamsout{\bgroup\markoverwith{\textcolor{violet}{\rule[0.5ex]{2pt}{1pt}}}\ULon}

\newcommand{\Ker}{\operatorname{Ker}}

\newcommand{\Rank}{\operatorname{Rank}}

\newcommand{\real}{\mathbb{R}}

\newcommand{\T}{\mathsf{T}} 

\newcommand{\mc}{\mathcal}

\newcommand{\sts}{\text{ss}}

\newcommand{\s}{\text{s}}
\newcommand{\Null}{\text{null}}

\DeclareSymbolFont{bbold}{U}{bbold}{m}{n}
\DeclareSymbolFontAlphabet{\mathbbold}{bbold}

\newcommand\oprocendsymbol{\hbox{$\square$}}
\newcommand\oprocend{\relax\ifmmode\else\unskip\hfill\fi\oprocendsymbol}

\renewcommand{\theenumi}{(\roman{enumi})}

\newcommand*{\QEDA}{\hfill\ensuremath{\blacksquare}}%

\graphicspath{{figs/}}

\makeatletter
\let\NAT@parse\undefined
\makeatother
\usepackage[colorlinks,urlcolor=blue, linkcolor=blue, citecolor=blue]{hyperref}

\def\BibTeX{{\rm B\kern-.05em{\sc i\kern-.025em b}\kern-.08em
    T\kern-.1667em\lower.7ex\hbox{E}\kern-.125emX}}

\begin{document}

\title{Model-based and Data-based Dynamic Output Feedback for Externally Positive Systems}

\author{Abed~AlRahman~Al~Makdah and Fabio~Pasqualetti
\thanks{This material is based upon work supported in part by awards
    ONR-N00014-19-1-2264.  A. A. Al Makdah
    and F. Pasqualetti are with the Department of Electrical and
    Computer Engineering and the Department of Mechanical Engineering
    at the University of California, Riverside, respectively,
    \href{mailto:aalmakdah@engr.ucr.edu}{\{\texttt{aalmakdah}},\href{mailto:fabiopas@engr.ucr.edu}{\texttt{fabiopas\}@engr.ucr.edu}}.}}
    
%

\maketitle

\begin{abstract}
In this work, we derive dynamic output-feedback controllers that render the closed-loop system externally positive. We begin by expressing the class of discrete-time, linear, time-invariant systems and the class of dynamic controllers in the space of input-output behaviors, where a dynamic controller can be expressed as a static behavioral feedback gain. We leverage the static form of the controller to derive output-feedback controllers that achieve monotonic output tracking of a constant non-negative reference output. Further, we provide a direct data-driven approach to derive monotonic tracking output-feedback controllers for single-input-single-output (SISO) systems.  Our approaches, model-based and data-based, allow us to obtain output-feedback controllers that render the closed-loop system externally positive. Finally, we validate our results numerically in a drone landing control~problem.
\end{abstract}


\section{Introduction}\label{sec:introduction}
In many natural and technological phenomena, the variables of interest are physical quantities that are naturally non-negative (e.g., population, concentration, charge levels, light intensity, prices, etc.). Mathematical models that are used to describe the dynamics of such variables should incorporate the non-negativity constraint. This motivates the study of a class of systems called \emph{positive systems}, which are systems whose states and output have non-negative evolution under any non-negative input and non-negative initial state~\cite{TK:02}. Positive systems have been used in several research areas, such as biology and pharmacology \cite{WMH-VC-QH:10,JAJ:72}, chemical reaction systems \cite{MF:95}, economics \cite{JWN-CIB-AL:86}, power systems \cite{AZ-TC-FK:12}, traffic and congestion \cite{RS-FW-DL:06}, and Markov chains and stochastic models \cite{ES:06}. The class of \emph{externally positive systems} is a relaxed class of positive systems, where a system is said to be~externally positive if its output is non-negative for any non-negative input and zero initial state~\cite{TK:02}. In this work, we address the problem of output-feedback control design that renders a given system externally positive. Although, as discussed below, this problem has been solved for specific cases (e.g., SISO systems and systems starting from rest), to the best of our knowledge, no general solution to this problem exists~\cite{AS-JL:20}.\\
\textbf{Related work.} The literature is rich with the analysis of positive systems \cite{TK:02,LF-SR:00}, where researchers studied the properties of positive systems, such as reachability, controllability, and observability. A stream of research focuses on feedback control synthesis for positive systems. In \cite{VGR-DJGJ:95}, the authors study the problem of pole-assignment for SISO positive systems. In \cite{TK:99}, the author provides sufficient conditions for the existence of stabilizing state-feedback controllers that ensure closed-loop positivity using Gershgorin's theorem. In \cite{HG-JL-CW-SX:05}, the authors provide necessary and sufficient conditions for the existence of state-feedback controllers that guarantee closed-loop positivity and asymptotic stability. In \cite{BR-EJD:09,MAR:11b}, the authors address the problem of designing static output-feedback controllers that ensure closed-loop positivity and asymptotic stability for SISO systems. Although in \cite{MAR:11b} the author investigates the MIMO case, the rank constraint on the controller gain matrix adds a limitation on the applicability of the proposed method. In \cite{BS-AM-MS:22}, the authors propose a direct data-driven approach to solve for stabilizing state-feedback controllers that ensure closed-loop positivity. We refer the reader to \cite{AR-MEV:18} for a more detailed list of references. Unlike the class of positive systems, fewer research has been invested in developing design techniques for controllers that ensure closed-loop external positivity. External positive systems have non-negative impulse response \cite{TK:02}. Several control design approaches focused on designing controllers that render the closed-loop impulse response non-negative \cite{SJ-JWS:92,SD:03,HT-MJ:20,HT-MJ:21}, which is equivalent to achieving a monotonic closed-loop step response. In \cite{SJ-JWS:92}, the authors present an approach to design compensators that achieve non-overshooting closed-loop response, which is based on pole-zero placement. In \cite{SD:03}, the author provides a compensator structure that ensures non-negative impulse response. In \cite{HT-MJ:20,HT-MJ:21}, the authors formulate linear programming approach for designing output-feedback controllers that ensure asymptotic stability and monotonic step-response. These approaches assume SISO systems and systems starting from rest. In \cite{RS-LN:12,EG-LN:15}, the authors provide approaches to design monotonic tracking state-feedback controllers for any initial state for MIMO systems. In \cite{CG-AR:21}, the authors provide a sufficient certificate for external positivity, which they use to design state-feedback controllers that ensure closed-loop external positivity. In general, the problem of designing an output-feedback controller that renders the closed-loop system externally positive is an open problem \cite{AS-JL:20}.\\
\textbf{Contributions.} This note features two contributions. First, we leverage the approach in \cite{AAAM-VK-VK-FP:22} to write an equivalent representation of discrete-time, linear, time-invariant systems, and dynamic output-feedback controllers in the behavioral space (Section \ref{sec: behavioral representation}), where we express dynamic output-feedback controllers as static behavioral feedback controllers in the behavioral space. Then, we use the system and controller's behavioral representations along with the formulation in \cite{EG-LN:15} to derive dynamic output-feedback controllers that monotonically track a constant non-negative reference output starting from a non-negative initial output (Section \ref{sec: monotonic control}). Second, we provide a direct data-driven approach to design dynamic output-feedback controllers that monotonically track a constant non-negative reference output starting from non-negative initial output for SISO systems (Section \ref{sec: data-driven monotonic control}), where we use input-output data collected from a system with unknown dynamics. Both contributions allow us to design output-feedback controllers (model-based or data-based) that render the closed-loop system externally positive. The data-driven approach makes it amenable to directly design output-feedback controllers online using data collected from a single input-output trajectory without the need to identify the system. We verify our theoretical results numerically for a drone landing control problem (Section~\ref{sec: drone landing}).

\section{Problem formulation}
Consider the discrete-time, linear, time-invariant system
\begin{align}
\begin{aligned}\label{eq: system in x}
  x(t+1) &= Ax(t) + Bu(t), \quad t\geq 0, \\ 
  y(t) &= Cx(t), 
 \end{aligned}
\end{align}
where $x(t)\in\real^{n}$ denotes the state, $u(t)\in \real^{m}$ the
control input, $y(t)\in\real^{p}$ the measured output, and the system matrices $A$, $B$ and $C$ have appropriate dimensions. For~the system \eqref{eq: system in x}, the output-feedback controller is written in~the~form:
\begin{align}\label{eq: output feedback controller}
\begin{split}
 x_c(t+1)&=Ex_c(t)+Fy(t), \quad t\geq 0\\
 u(t)&=Gx_c(t)
 \end{split}
\end{align}
where $x_c(t) \in \mathbb{R}^n$ denotes the controller's internal state at time $t$, and the matrices $E$, $F$, and $G$ have appropriate dimensions. The output feedback controller using Luenberger observer can be written in the form \eqref{eq: output feedback controller} with $E=A-BK-LC$, $F=L$, and $G=-K$, where $K \in \mathbb{R}^{m\times n}$ and $L\in \mathbb{R}^{n\times p}$ are the controller gain and the observer gain, respectively. Throughout this note, we adopt the following notation, for a vector $w\in \mathbb{R}^q$, $w \geq 0$ implies that all the components of $w$ are non-negative. Before we state our problem formulation, we introduce the following definition from the literature.
%
\begin{definition}{\bf \emph{(Externally positive system \cite{TK:02})}}\label{def: external positive system}
The system \eqref{eq: system in x} is called externally positive if and only if for every input sequence $u(t)\geq 0$ for $t\geq 0$, and $x(0)=0$, the output $y(t)\geq 0$ for $t\geq0$. \oprocend
\end{definition}
This paper focuses~on designing an output feedback controller that renders the~closed-loop system to be externally positive, i.e., the closed-loop output response of \eqref{eq: system in x} is non-negative $\forall t\geq 0$. Formally:
%
\begin{problem}\label{pr: problem}
Given a system \eqref{eq: system in x} with relative degree $d$, design a controller in the form \eqref{eq: output feedback controller} such that for any $y(0) \geq 0$:
\begin{enumerate}
\item The closed-loop system is asymptotically stable.
\item The closed-loop output response converges to a desired non-negative output $y_{\sts}$.
\item The closed-loop system is externally positive.~\oprocend
\end{enumerate}
\end{problem} 
%
For Problem \ref{pr: problem} to be solvable, the following condition is required.
\begin{assumption}{\bf \emph{(Assumption on the initial output)}}\label{assump: initial output}
Given a system \eqref{eq: system in x} with relative degree $d$ and any $y(0)\geq 0$, the output $\{y(1),\cdots , y(d-1)\}$ is non-negative.~\oprocend
\end{assumption}
\smallskip
Notice that, the sequence $\{y(1),\cdots , y(d-1)\}$ cannot be affected by any input because of the relative degree $d$. Hence, Assumption \ref{assump: initial output} is necessary for the solvability of Problem~\ref{pr: problem}.\\
In this note, we tackle Problem \ref{pr: problem} in model-based and data-based settings, where we focus on designing monotonic tracking controllers to track a constant non-negative reference output starting from non-negative initial output. The monotonic tracking ensures that the closed-loop output converges to a non-negative reference output without overshooting nor undershooting, which guarantee that the closed-loop output remains non-negative for all non-negative initial outputs. Thus, solving Problem \ref{pr: problem} boils down to designing a monotonic tracking dynamic output-feedback controller. To this aim, first in Section~\ref{sec: behavioral representation}, we leverage the approach in \cite{AAAM-VK-VK-FP:22} to provide an equivalent representation of \eqref{eq: system in x} and \eqref{eq: output feedback controller} in the states of the system's input-output behaviors, where the form of \eqref{eq: output feedback controller} is converted into a form of static feedback controller. Then in Section \ref{sec: monotonic control}, we use the behavioral system and controller representations from Section~\ref{sec: behavioral representation} and the results in \cite{EG-LN:15} to design a behavioral feedback controller  that monotonically tracks a constant non-negative reference output. In Section~\ref{sec: data-driven monotonic control}, we provide a direct data-driven approach to design a behavioral feedback monotonic tracking controller for SISO systems using input-output data collected from one experiment. The following standard assumptions on system \eqref{eq: system in x} are required to ensure tracking of a constant reference output for any initial condition \cite{EG-LN:15}.
\begin{assumption}{\bf \emph{(Assumptions on system \eqref{eq: system in x})}}\label{assump: reference tracking}
The pairs $(A,B)$ and $(A,C)$ are stabilizable and observable, respectively, and system \eqref{eq: system in x} is right-invertible and has no invariant zeros equal to $1$.~\oprocend
\end{assumption}
\section{Behavioral representation}\label{sec: behavioral representation}
In this section, we derive an equivalent representation of the system \eqref{eq: system in x} and the controller \eqref{eq: output feedback controller} in the space of input-output behaviors. To this aim, we define the behavioral space of~\eqref{eq: system in x}~as
\setcounter{equation}{2}
\begin{align}\label{eq: z}
  z(t)\triangleq[U(t-1)^{\T},Y(t-1)^{\T}]^{\T} ,
\end{align}
where
\begin{align}
  U(t-1)&\triangleq\left[u(t-n)^{\T}, \cdots, u(t-1)^{\T}\right]^{\T}, \nonumber\\
  Y(t-1)&\triangleq\left[y(t-n)^{\T}, \cdots, y(t-1)^{\T}\right]^{\T}. \nonumber
  \end{align}
By leveraging the approach in \cite{AAAM-VK-VK-FP:22}, we can write \eqref{eq: system in x} in the behavioral space $z$ as
\begin{align} \label{eq: z dynamics}
\begin{footnotesize}
\begin{aligned}
\underbrace{\left[ \begin{smallarray}{c}
u(t-n+1) \\
\vdots \\
u(t-1)\\
u(t) \\ 
\hdashline[2pt/0pt] 
y(t-n+1)\\
\vdots \\
y(t-1)\\
y(t)
\end{smallarray} \right]}_{z(t+1)}=&
\underbrace{\left[ \begin{smallarray}{ccccc;{2pt/0pt}ccccc}
0 & I & 0 & \cdots & 0 & 0 & 0 & 0& \cdots & 0\\
\vdots & \vdots & \vdots & \ddots & \vdots & \vdots & \vdots & \vdots & \ddots & \vdots \\
0 & 0 & 0 & \cdots & I & 0 & 0 & 0& \cdots & 0\\
0 & 0 & 0 & \cdots & 0 & 0 & 0 & 0& \cdots & 0\\
\hdashline[2pt/0pt] 
0 & 0 & 0 & \cdots & 0 & 0 & I & 0 & \cdots & 0 \\
\vdots & \vdots & \vdots & \ddots & \vdots & \vdots & \vdots & \vdots & \ddots & \vdots  \\
0 & 0 & 0& \cdots  & 0 & 0 & 0 & 0 & \cdots & I \\
\multicolumn{5}{c;{2pt/0pt}}{\mc{A}_u} & \multicolumn{5}{c}{\mc{A}_y}
\end{smallarray} \right]}_{\mc{A}}
\underbrace{\left[ \begin{smallarray}{c}
u(t-n) \\
\vdots \\
u(t-2) \\
u(t-1) \\ 
\hdashline[2pt/0pt] 
y(t-n)\\
\vdots \\
y(t-2)\\
y(t-1)
\end{smallarray} \right]}_{z(t)}
+\underbrace{\left[ \begin{smallarray}{c}
0\\
\vdots\\
0 \\
I   \\
\hdashline[2pt/0pt] 
0 \\
\vdots \\
0 \\
0 \\ 
\end{smallarray} \right]}_{\mc{B}}
u(t),\\
y_z(t)=&\underbrace{\left[\begin{smallarray}{ccc;{2pt/0pt}cccc}
0 & \cdots & 0 & 0 &  \cdots & 0 & I
\end{smallarray}\right]}_{\mc{C}} z(t).
\end{aligned}
\end{footnotesize}
\end{align}
We refer the reader to Appendix~\ref{app: behavioral dynamics} for the derivation of~\eqref{eq: z dynamics}. This implies that given a sequence of control inputs, the state $z$ contains the system output $y$ over time, and can be used to reconstruct the exact value of the system state~$x$. This also implies that a controller for the system \eqref{eq: system in x} can~equivalently be designed using the dynamics \eqref{eq: z dynamics}. In \cite[Lemma 5.3]{AAAM-VK-VK-FP:22}, the authors show that any dynamic controller \eqref{eq: output feedback controller} for \eqref{eq: system in x} can be equivalently represented as a static controller for~\eqref{eq: z dynamics}
%
%
\setcounter{equation}{4}
\begin{align}\label{eq: static controller}
u(t)=\mc{K}z(t),
\end{align}
where $\mc{K}\in \mathbb{R}^{m \times r}$ is a static feedback gain and $r=n(m+p)$. This implies that designing an output feedback controller in the form of \eqref{eq: output feedback controller} for the system \eqref{eq: system in x} is equivalent to designing a static-feedback controller in the form of \eqref{eq: static controller} for the system \eqref{eq: z dynamics}. We can equivalently rewrite Problem \ref{pr: problem} as
\begin{problem}\label{pr: problem z}
Given a system \eqref{eq: z dynamics} with relative degree $d$, design a controller in the form \eqref{eq: static controller} such that for any $\left. y_z(n+d)\geq 0\right.$:
\begin{enumerate}
\item The closed-loop system is asymptotically stable.
\item The closed-loop output response converges to a desired non-negative output $y_{\sts}$.
\item The closed-loop system is externally positive.~\oprocend
\end{enumerate}
\end{problem} 
%
The following result provides a necessary and sufficient condition under which Problem \ref{pr: problem z} is equivalent to Problem~\ref{pr: problem}.
%
%
\begin{theorem}{\bf \emph{(Equivalence of Problem \ref{pr: problem} and \ref{pr: problem z})}}\label{thrm: equivalence} Problem \ref{pr: problem z} is equivalent to Problem \ref{pr: problem} under Assumption \ref{assump: initial output} if and only if there exist an input sequence $\left.\{u(0),\cdots, u(n-1)\}\right.$ such that the corresponding output sequence $\{y(d),\cdots, y(n+d-1)\}$ is non-negative.
\begin{IEEEproof}
(\emph{Sufficiency}) Let Assumption \ref{assump: initial output} holds, and let $\left.\{u(0),\cdots, u(n-1)\}\right.$ such that the corresponding~output $\{y(d),\cdots, y(n+d-1)\}$ is non-negative. Since~the~systems \eqref{eq: system in x} and \eqref{eq: z dynamics} are equivalent (by Lemma~\ref{lemma: system in z}), and the controllers \eqref{eq: output feedback controller} and \eqref{eq: static controller} are equivalent (by \cite[Lemma 5.3]{AAAM-VK-VK-FP:22}), then Problem \ref{pr: problem z} and Problem \ref{pr: problem} are equivalent.\\
\indent(\emph{Necessity}) We show this via contrapositive. Let Assumption \ref{assump: initial output} be satisfied and assume that a sequence $\left.\{u(0),\cdots, u(n-1)\}\right.$ such that the corresponding output sequence $\{y(d),\cdots, y(n+d-1)\}$ is non-negative does not exist. Then, a solution to Problem \ref{pr: problem} does not exist, while a solution to Problem \ref{pr: problem z} exists. Hence, Problem \ref{pr: problem} and Problem \ref{pr: problem z} are not equivalent.
\end{IEEEproof}
\end{theorem}
%
We verify the condition in Theorem \ref{thrm: equivalence} in Appendix \ref{app: equivalence}.
\section{Model-based monotonic tracking control}\label{sec: monotonic control}
In this section, we design a controller in the form \eqref{eq: static controller} that solves Problem \ref{pr: problem z}, and equivalently solves Problem \ref{pr: problem} under Assumption \ref{assump: initial output} and the condition in Theorem \ref{thrm: equivalence}. To this aim, we design a monotonic tracking controller~for system \eqref{eq: z dynamics} to track a constant non-negative reference output~$y_{\sts}$ for any $\left. y_z(d+n)\geq 0\right.$.
Assumption \ref{assump: reference tracking} guarantees the existence of $x_{\text{ss}} \in \mathbb{R}^n$ and $u_{\text{ss}} \in \mathbb{R}^m$ that~satisfy
\begin{align*}
\begin{split}
x_{\sts}&=A x_{\sts}+ B u_{\sts},\\
y_{\sts}&=C x_{\sts},
\end{split}
\end{align*}
for any $y_{\sts}\in \mathbb{R}^p$. Equivalently, via Lemma \ref{lemma: system in z}, we can write
\begin{align*}
\begin{split}
z_{\sts}&=\mc{A} z_{\sts}+ \mc{B} u_{\sts},\\
y_{\sts}&=\mc{C} z_{\sts},
\end{split}
\end{align*}
where $z_{\sts}=\left[u_{\sts}^{\T}, \cdots, u_{\sts}^{\T}, y_{\sts}^{\T}, \cdots, y_{\sts}^{\T}\right]^{\T}$. Let $\zeta(t)\triangleq z(t)-z_{\sts}$ and $\epsilon(t)\triangleq y_z(t)-y_{\sts}$. Then, we can write the dynamics of $\zeta$ and $\epsilon$ as
\begin{align}\label{eq: zeta dynamics}
\begin{split}
 \zeta(t+1)&=\mc{A} \zeta(t) + \mc{B} v(t), \qquad t\geq 0,\\
 \epsilon(t)&=\mc{C} \zeta(t),
\end{split}
\end{align}
where $v(t)=\mc{K} \zeta(t)$. The tracking controller applied to \eqref{eq: z dynamics} takes the form
\begin{align}\label{eq: tracking control}
u(t)=\mc{K} \left(z(t)-z_{\sts}\right)+ u_{\sts}= v(t)+u_{\sts}.
\end{align}
Note that $\epsilon(t)$ in \eqref{eq: zeta dynamics} can be either negative ($y(t)<y_{\sts}$) or positive ($y(t)>y_{\sts}$) even when both $y(t)$ and $y_{\sts}$ are non-negative. Therefore, we need each entry of $\epsilon(t)$ to converge monotonically to zero. According to \cite{EG-LN:15}, global monotonicity can be obtained if and only~if
\begin{align}\label{eq: output error evolution}
 \epsilon(t)=\begin{bmatrix}
 \alpha_1 \lambda_1^t & \cdots & \alpha_p \lambda^t_p 
 \end{bmatrix}^{\T},
\end{align}
where $\{\lambda_1,  \dots, \lambda_p\}$ are positive real and less than $1$, and $\{\alpha_1 , \dots, \alpha_p\}$ are real constants that depend on the initial conditions. The following Theorem is adapted for our setting from \cite[Theorem 3.1]{EG-LN:15} and it provides necessary and sufficient conditions for the existence of a controller that solves Problem~\ref{pr: problem z}.
%
\begin{theorem}{\bf \emph{(Necessary and sufficient conditions for the solvability of Problem \ref{pr: problem z})}}\label{thrm: nec and suff cond}
 Let $\lambda_1 , \dots , \lambda_p \in [0,1)$. There exists a feedback gain $\mc{K} \in \mathbb{R}^{m \times r}$ that solves Problem \ref{pr: problem z} if and only if there exist $M \in \mathbb{R}^{r \times r}\succ 0$ and $N \in \mathbb{R}^{m\times r}$ that solve the following LMI problem:
\begin{align}
&\begin{bmatrix}
M & \mc{A} M + \mc{B} N \\
M^{\T}\mc{A}^{\T} + N^{\T} \mc{B}^{\T} & M
\end{bmatrix} \succ 0, \label{eq: LMI ineq}\\
&~\mc{C}_i \left(\mc{A} M+\mc{B} N\right)=\lambda_i \mc{C}_i N, \qquad i \in {1, \dots , p}. \label{eq: LMI eq}
\end{align}
 Further, the gain $\mc{K}=N M^{-1}$ solves Problem \ref{pr: problem z}.
\begin{IEEEproof}
The closed-loop stability of \eqref{eq: zeta dynamics} with $v=\mc{K}\zeta$ is guaranteed if and only if there exist $P \in \mathbb{R}^{r\times r} \succ 0$ that satisfies the following Lyapunov inequality
\begin{align}\label{eq: lyap inequality}
\left(\mc{A} + \mc{B} \mc{K} \right) P \left(\mc{A} + \mc{B} \mc{K} \right)^{\T} -P \prec 0.
\end{align}
Let $M=P$ and $\mc{K}=N M^{-1}$, then we can re-write \eqref{eq: lyap inequality} as
\begin{align}\label{eq: lyap inequality rewritten}
M-\left(\mc{A}M + \mc{B}N \right) M^{-1} \left(\mc{A}M + \mc{B}N \right)^{\T}  \succeq 0.
\end{align}
The condition \eqref{eq: LMI ineq} is satisfied if and only if \eqref{eq: lyap inequality rewritten} is satisfied, this can be observed by taking the Schur complement of \eqref{eq: LMI ineq} with respect to the $(2,2)$-block. From \cite[Lemma 3.1]{EG-LN:15}, the closed-loop output tracking error is in the form of \eqref{eq: output error evolution} for $\lambda_1, \cdots \lambda_p \in [0,1)$ if and only if there exist $\mc{K}\in \mathbb{R}^{m \times r}$ such~that
\begin{align}\label{eq: monotonicity condition}
 \mc{C}_i \left(\mc{A} + \mc{B} \mc{K} \right) = \lambda_i \mc{C}_i,
\end{align}
where $\mc{C}_i$ is the $i$-th row of $\mc{C}$ for $i\in\{1,\cdots,p\}$. The condition \eqref{eq: LMI eq} follows from \eqref{eq: monotonicity condition} by multiplying $M$ from the right. Conditions \eqref{eq: LMI ineq} and \eqref{eq: LMI eq} are necessary and sufficient for achieving stability and global monotonicity for the system \eqref{eq: zeta dynamics} with $v=NM^{-1}\zeta$ for all initial conditions. Hence, conditions \eqref{eq: LMI ineq} and \eqref{eq: LMI eq} are sufficient for the solvability of Problem \ref{pr: problem z}. To show the necessity, let $v(t)$ for $t\geq n$ be a controller such that condition \eqref{eq: LMI eq} is not satisfied, then starting from any initial state such that $y_z(d+n)\geq 0$ and under controller $v(t)$, the tracking error $\epsilon(t)$ will not satisfy \eqref{eq: output error evolution}, and hence the output will not monotonically track a constant non-negative reference output and might overshoot to negative values. Therefore, Problem 2 is not solvable. Similarly, if condition \eqref{eq: LMI ineq} is not satisfied under $v(t)$, the system \eqref{eq: zeta dynamics} will be unstable and hence Problem \ref{pr: problem z} is unsolvable.
%
\end{IEEEproof}
\end{theorem}
%
%
Note that condition \eqref{eq: LMI ineq} ensures that the obtained controller is stabilizing, and condition \eqref{eq: LMI eq} ensures that $\epsilon$ is in the form \eqref{eq: output error evolution} and hence ensures the monotonicity of the closed-loop output response.
%
%
\section{Data-based monotonic tracking control}\label{sec: data-driven monotonic control}
In this section, we design a controller in the form \eqref{eq: static controller} that solves Problem \ref{pr: problem z} using input-output data collected from system \eqref{eq: system in x} using one experiment. Throughout this section, we assume that system \eqref{eq: system in x} is a single-input-single-output (SISO) system (i.e., $m=p=1$), and the steady-state output and steady-state input equal to zero (i.e., $y_{\sts}=0$ and $u_{\sts}=0$).{\footnote{Setting $y_{\sts}=0$ and $u_{\sts}=0$ does not affect the generality of our results (see Remark \ref{rmrk: tracking}).}} We use the following notation to express the data collected from a trajectory with time horizon~$T$:
\begin{align}\label{eq: input output data}
\begin{split}
 u_{0:T}\triangleq \left[u(0),\cdots , u(T)\right], ~~
 y_{0:T}\triangleq \left[y(0),\cdots , y(T)\right],
 \end{split}
\end{align}
with the corresponding Hankel matrix{\footnote{We adopt similar notation as \cite{CDP-PT:19}: the first subscript denotes the time at which the first sample is taken, and the second and the third subscripts denote the number of samples per each column and row,~respectively.}}
\begin{align}\label{eq: hankel matrix}
 \begin{bmatrix}
 U_{0,n,T-n+1}\\
 \hdashline[2pt/0pt] 
 Y_{0,n,T-n+1}
 \end{bmatrix}
 \triangleq 
\left[\begin{smallarray}{ccc}
 u(0)    & \cdots & u(T-n)\\
 \vdots & \ddots & \vdots \\
 u(n-1) & \cdots & u(T-1) \\
  \hdashline[2pt/0pt] 
   y(0)    & \cdots & y(T-n)\\
 \vdots & \ddots & \vdots \\
y(n-1) & \cdots & y(T-1)
\end{smallarray}\right].
\end{align}
Notice that the input-state data for system \eqref{eq: z dynamics} can be obtained from the input-output data in \eqref{eq: input output data} and \eqref{eq: hankel matrix} as
\begin{align}\label{eq: behavioral data}
\begin{split}
 u_{n:T}\triangleq \left[u(n),\cdots , u(T)\right],\quad
 z_{n:T}\triangleq \begin{bmatrix}
 U_{0,n,T-n+1}\\
 \hdashline[2pt/0pt] 
 Y_{0,n,T-n+1}
 \end{bmatrix}.
\end{split}
\end{align}
%
%
Next, we make use of the results in \cite{CDP-PT:19} to get a data-dependent representation of the closed-loop dynamics of~system \eqref{eq: z dynamics}. Then, we derive conditions equivalent to the conditions in Theorem \ref{thrm: nec and suff cond} in terms of the data in \eqref{eq: behavioral data}. In what follows, we require the following Assumption and Lemma.
\smallskip
\begin{assumption}{\bf \emph{(Persistency of excitation)}}\label{assump: persistency of excitation}
The data in \eqref{eq: input output data} is collected with an input, $u_{0:T}$ with $T\geq 4n$, persistently exciting of order $2n+1$.~\oprocend
\end{assumption}
%
%
\begin{lemma}{\bf \emph{(Rank condition)}}\label{lemma: rank condition}
Given data as in \eqref{eq: behavioral data} collected from a SISO system \eqref{eq: system in x} with an input, $u_{0:T}$, that satisfies Assumption \ref{assump: persistency of excitation}. Then
\begin{align}\label{eq: rank condition}
\Rank\left(\begin{bmatrix}
u_{n:T} \\
\hdashline[2pt/0pt]
z_{n:T}
\end{bmatrix}\right)=2n+1.
\end{align}
\begin{IEEEproof}
We start by noting that 
\begin{align*}
\Rank\left(\begin{bmatrix}
u_{n:T} \\
\hdashline[2pt/0pt]
z_{n:T}
\end{bmatrix}\right)
&= \Rank\left(\begin{bmatrix}
u_{n:T} \\
\hdashline[2pt/0pt]
 U_{0,n,T-n+1}\\
 \hdashline[2pt/2pt] 
 Y_{0,n,T-n+1}
\end{bmatrix}\right)\\
&= \Rank\left(\begin{bmatrix}
U_{0,n+1,T-n+1}\\
 \hdashline[2pt/2pt] 
Y_{0,n,T-n+1}
\end{bmatrix}\right).
\end{align*}
Using \eqref{eq: system in x}, we can write the input-output response as
\begin{align}\label{eq: input-output response}
 \begin{bmatrix}
U_{0,n+1,T-n+1}\\
 \hdashline[2pt/2pt] 
Y_{0,n,T-n+1}
\end{bmatrix} 
=
\underbrace{\begin{bmatrix}
 I & 0 \\
\mc{F} & \mc{O}
\end{bmatrix}}_{\triangleq \mc{H}}
\begin{bmatrix}
U_{0,n+1,T-n+1}\\
 x_{0:T-n} 
\end{bmatrix},
\end{align}
where
\begin{align*}
\mc{O}&\triangleq \left[\begin{smallarray}{c}
      C\\
      CA\\
      \vdots \\
      CA^{n-1}
    \end{smallarray}\right], \quad 
    \mc{F}\triangleq\left[\begin{smallarray}{ccccc}
      0    & \cdots & 0 & 0 & 0\\
      CB & \cdots & 0 & 0 & 0 \\
      \vdots  & \ddots & \vdots & \vdots & \vdots \\
      CA^{n-2}B & \cdots & CB & 0 & 0
    \end{smallarray}\right].
\end{align*}
For SISO systems, the matrix $\mc{H}$
is square and full-rank with $\Rank{\left(\mc{H}\right)} = 2n+1$, and since the input signal $u_{0:T}$ satisfies Assumption \ref{assump: persistency of excitation}, using \cite[Corollary 2-(iii)]{JCW-PR-IM-BLMDM:05} we~have 
\begin{align*}
\Rank\left(\begin{bmatrix}
U_{0,n+1,T-n+1}\\
 x_{0:T-n} 
\end{bmatrix}\right) = 2n+1,
\end{align*}
then we have 
\begin{align*}
\Rank\left(\begin{bmatrix}
u_{n:T} \\
\hdashline[2pt/0pt]
z_{n:T}
\end{bmatrix}\right)
&= \Rank\left(\begin{bmatrix}
U_{0,n+1,T-n+1}\\
 \hdashline[2pt/2pt] 
Y_{0,n,T-n+1}
\end{bmatrix}\right)=2n+1.
\end{align*}
\end{IEEEproof}
\end{lemma}
%
%
The next Theorem provides necessary and sufficient conditions equivalent to the conditions in Theorem \ref{thrm: nec and suff cond} in terms of the data in~\eqref{eq: behavioral data}.
\begin{theorem}{\bf \emph{(Necessary and sufficient conditions for the solvability of Problem \ref{pr: problem z} using data)}}\label{thrm: nec and suff cond data}
Let the data in \eqref{eq: input output data} be collected from a SISO system \eqref{eq: system in x} with the input $u_{0:T}$ satisfying Assumption \ref{assump: persistency of excitation}. Let $\lambda \in [0,1)$. There exists a feedback gain $\mc{K} \in \mathbb{R}^{m \times r}$ that solves Problem \ref{pr: problem z} if and only if there exist $P \in \mathbb{R}^{r \times r}\succ 0$ and $Q \in \mathbb{R}^{m\times r}$ that solve the following LMI problem:
\begin{align}
& \begin{bmatrix}
P & z_{n+1:T+1}Q \\
Q^{\T}z_{n+1:T+1}^{\T} & P
\end{bmatrix} \succ 0,\label{eq: LMI data-ineq}\\
&\qquad \quad z_{n:T} Q =P,\label{eq: LMI data-eq1}\\
&~\mc{C} z_{n+1:T+1} Q =\lambda \mc{C} P.\label{eq: LMI data-eq2}
\end{align}
 Further, the gain $\mc{K}=u_{n:T}Q P^{-1}$ solves Problem \ref{pr: problem z}.
\begin{IEEEproof}
In the proof of Theorem \ref{thrm: nec and suff cond}, we show that the controller $u=\mc{K} z$ solves Problem \ref{pr: problem z} if and only if there exist $P\succ 0$ such that \eqref{eq: lyap inequality} and \eqref{eq: monotonicity condition} are satisfied.
%
Next, we express conditions \eqref{eq: lyap inequality} and \eqref{eq: monotonicity condition} in terms of the data \eqref{eq: behavioral data}. Since $u_{0:T}$ satisfies Assumption \ref{assump: persistency of excitation}, the rank condition in \eqref{eq: rank condition} holds, then using \cite[Theorem 2]{CDP-PT:19}, we can write the closed-loop dynamics of \eqref{eq: z dynamics} as
\begin{align}\label{eq: closed-loop data}
 \mc{A} +\mc{B} \mc{K}= z_{n+1:T+1} \mc{G},
\end{align}
where $\mc{G} \in \mathbb{R}^{(T-n+1)\times r}$ satisfies 
\begin{align}\label{eq: G equation}
\begin{bmatrix}
 \mc{K}\\
 I_r
\end{bmatrix}
=
\begin{bmatrix}
 u_{n:T}\\
 z_{n:T}
\end{bmatrix} \mc{G}.
\end{align}
Let $Q \triangleq \mc{G} P$, then conditions \eqref{eq: lyap inequality} and \eqref{eq: monotonicity condition} are equivalent to the existence of $Q$ and $P\succ 0$ such that
\begin{align}
&z_{n+1:T+1} Q P^{-1} Q^{\T} z_{n+1:T+1}^{\T} - P \prec 0,\label{eq: LMI proof-ineq}\\
& z_{n:T} Q  = P, \label{eq: LMI proof-eq1}\\
& \mc{C} z_{n+1:T+1} Q = \lambda \mc{C} P, \label{eq: LMI proof-eq2}
\end{align}
where \eqref{eq: LMI proof-ineq} and \eqref{eq: LMI proof-eq2} are obtained by substituting \eqref{eq: closed-loop data} into \eqref{eq: lyap inequality} and \eqref{eq: monotonicity condition}, respectively, and \eqref{eq: LMI proof-eq1} is obtained from \eqref{eq: G equation}. The inequalities \eqref{eq: LMI data-ineq} and \eqref{eq: LMI proof-ineq} are equivalent, this can be observed by taking the Schur complement of \eqref{eq: LMI data-ineq} with respect to the $(2,2)$-block. Finally, from \eqref{eq: G equation}, we have $\mc{K}=u_{n:T} Q P^{-1}$.
\end{IEEEproof}
\end{theorem}
\smallskip
Note that conditions \eqref{eq: LMI data-ineq} and \eqref{eq: LMI data-eq1} ensure that the obtained controller is stabilizing, and they are equivalent to the~conditions derived in \cite[Theorem 8]{CDP-PT:19}. Condition \eqref{eq: LMI data-eq2}~ensures the monotonicity of the closed-loop output response,~where~$\lambda$~is~a design parameter that corresponds to the closed-loop eigenvalue that appears in \eqref{eq: output error evolution}, and $\mc{C}$ is the output matrix of the system in the behavioral space \eqref{eq: z dynamics}, which has a fixed structure. We conclude this section with the following remarks.
%
%
\begin{remark}{\bf \emph{(Non-zero steady state input and output)}}\label{rmrk: tracking}
When the steady-state input and output are non-zero, the controller takes the form \eqref{eq: tracking control}, where $\mc{K}$ is obtained as in Theorem \ref{thrm: nec and suff cond data} using the data collected in \eqref{eq: input output data}.~\oprocend
\end{remark}
%
%
\smallskip
\begin{remark}{\bf \emph{(MIMO case)}}\label{rmrk: MIMO case}
For the case of multi-input-multi-output (MIMO) systems, our result in Theorem \ref{thrm: nec and suff cond data} may not hold since the rank condition in \eqref{eq: rank condition} will not hold and the matrix $\begin{bmatrix}
u_{n:T} \\
\hdashline[2pt/0pt]
z_{n:T}
\end{bmatrix}$ loses rank, with $\Rank{\left(\begin{bmatrix}
u_{n:T} \\
\hdashline[2pt/0pt]
z_{n:T}
\end{bmatrix}\right)} \leq  (n+1)m+n$. This can be observed from \eqref{eq: input-output response}.~\oprocend
\end{remark}
%
%
%
%
%
\section {Drone landing control} \label{sec: drone landing}
In this section, we illustrate our theoretical results in a drone landing control problem. We consider drone dynamics that are restricted to the motion in the vertical axis and obey
%
\begin{align}\label{eq: drone dynamics}
\begin{split}
 x(t+1)&= 
\underbrace{\begin{bmatrix}
 1 & T_{\s}\\
 0 & 1
\end{bmatrix}}_{A} x(t)
+
\underbrace{\begin{bmatrix}
 0\\
 T_{\s}
\end{bmatrix}}_{B} u(t), \quad t\geq 0 ,\\
y(t)&= 
\underbrace{\begin{bmatrix}
 1 & 0
\end{bmatrix}}_{C}
x(t),
\end{split}
\end{align}
where $x(t) \in \mathbb{R}^2$ contains the drone's altitude and vertical velocity, $u(t) \in \mathbb{R}$ is the input signal, $y(t) \in \mathbb{R}$ is the output signal that corresponds to the drone's altitude, and $T_{\s}>0$ is the sampling time. In this example, we have $T_{\s}=0.1$. The ground level corresponds to $y=0$ and the elevation above the ground level corresponds to $y>0$. Our aim is to design an output feedback controller in the form \eqref{eq: tracking control}, which renders the drone to land safely without crashing into the ground, i.e., the closed-loop output should remain non-negative $\forall t\geq 0$ and $y(t)\rightarrow \!y_{\sts}$ as $t \rightarrow \!\infty$, with $u_{\sts}\!=\!0$ and $y_{\sts}\!=\!0$. We consider two settings: the model-based setting, where we have access to the system dynamics \eqref{eq: drone dynamics}, and the data-based setting, where we do not have access to the system dynamics but we have access to input-output data collected from \eqref{eq: drone dynamics} in the form~\eqref{eq: input output data}.
%
%
\subsubsection{Model-based}
\begin{figure}[!t]
  \centering
  \includegraphics[width=1\columnwidth,trim={0cm 0cm 0cm
    0cm},clip]{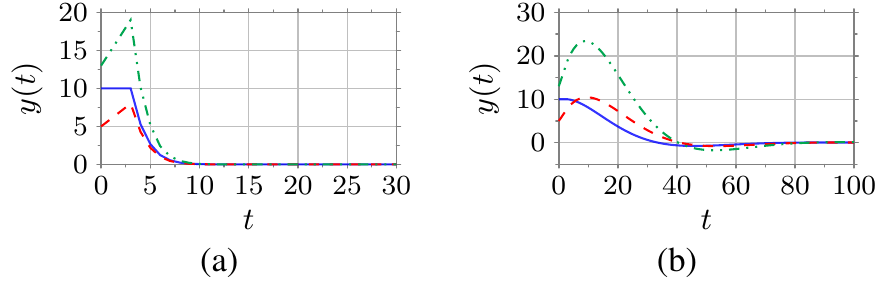}
  \caption{This figure shows the closed-loop output response of \eqref{eq: drone dynamics} driven by model-based controllers. In both panels, the solid blue line, dashed red line and the dash-dotted green line correspond to $x(0)=[10, 0]^{\T}$, $[5, 10]^{\T}$, and $[13, 20]^{\T}$, respectively. Panel (a) shows the closed-loop output response for the controller obtained using Theorem \ref{thrm: nec and suff cond}. Panel (b) shows the closed-loop output response for the controller obtained using Theorem \ref{thrm: nec and suff cond} but with relaxing condition \eqref{eq: LMI eq}, which ensures monotonicity of the closed-loop output response. Notice that Assumption \ref{assump: initial output} and Theorem \ref{thrm: equivalence} are satisfied.}
    \label{fig: drone_landing_model_based}
\end{figure}

Using Theorem \ref{thrm: nec and suff cond} with $\lambda=0.4$, we get $\mc{K}_1=\begin{bmatrix}-1.889 & -1.442 & 188.887 & -235.882\end{bmatrix}$. If we relax condition \eqref{eq: LMI eq} in Theorem \ref{thrm: nec and suff cond}, which ensures the monotonicity of the closed-loop output response, we obtain $\mc{K}_2=\begin{bmatrix} -0.317 & -0.464 & 18.571 & -19.785\end{bmatrix}$. Fig. \ref{fig: drone_landing_model_based} shows the closed-loop output response of system \eqref{eq: drone dynamics} driven by model-based controllers. Fig. \ref{fig: drone_landing_model_based}(a) shows the closed-loop output response for the controller $\mc{K}_1$ for different non-negative initial outputs. We observe that the output converges monotonically to the desired output without overshooting where the drone lands safely without crashing. Fig. \ref{fig: drone_landing_model_based}(b) shows the closed-loop output response for the controller $\mc{K}_2$ for the same initial conditions as in Fig. \ref{fig: drone_landing_model_based}(a). We observe that the output overshoots to negative values before converging to zero, which implies that the drone crashes into the ground.
\smallskip
\subsubsection{Data-based}
\begin{figure}[!t]
  \centering
  \includegraphics[width=1\columnwidth,trim={0cm 0cm 0cm
    0cm},clip]{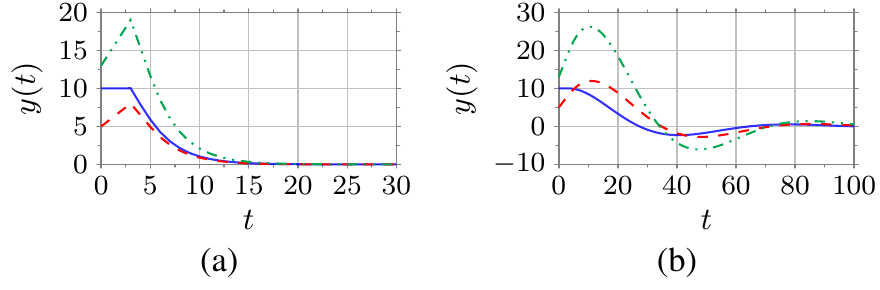}
  \caption{This figure shows the closed-loop output response of \eqref{eq: drone dynamics} driven by data-based controllers. In both panels, the solid blue line, dashed red line and the dash-dotted green line correspond to $x(0)=[10, 0]^{\T}$, $[5, 10]^{\T}$, and $[13, 20]^{\T}$, respectively. Panel (a) shows the closed-loop output response for the controller obtained using Theorem \ref{thrm: nec and suff cond data}. Panel (b) shows the closed-loop output response for the controller obtained using Theorem \ref{thrm: nec and suff cond data} but with relaxing condition \eqref{eq: LMI data-eq2}, which ensures monotonicity of the closed-loop output response. Notice that Assumption \ref{assump: initial output} and Theorem \ref{thrm: equivalence} are satisfied.}
    \label{fig: drone_landing_data_based}
\end{figure}

We collect data as in \eqref{eq: input output data} with the input, $u_{0:T}$, satisfying Assumption \ref{assump: persistency of excitation}. 
Using Theorem \ref{thrm: nec and suff cond data} with $\lambda=0.4$, we get $\mc{K}_3=\begin{bmatrix}-1.248 & -1.084 & 124.835 & -146.322\end{bmatrix}$. If we relax condition \eqref{eq: LMI data-eq2} in Theorem \ref{thrm: nec and suff cond data}, which ensures monotonicity of the closed-loop output response, we obtain $\mc{K}_4=\begin{bmatrix} 0.019 & 0.257 & 5.774 & -6.258\end{bmatrix}$. Fig. \ref{fig: drone_landing_data_based} shows the closed-loop output response of system \eqref{eq: drone dynamics} driven by data-based controllers. Fig. \ref{fig: drone_landing_data_based}(a) shows the closed-loop output response for the controller $\mc{K}_3$ for different non-negative initial outputs. We observe that the output converge monotonically to the desired output without overshooting where the drone lands safely without crashing. Fig. \ref{fig: drone_landing_data_based}(b) shows the closed-loop output response for the controller $\mc{K}_4$ for the same initial conditions as in Fig. \ref{fig: drone_landing_data_based}(a). We observe that the output overshoots to negative values before converging to zero, which implies that the drone crashes into the ground.
\section{Conclusion and future work} 
In this work, we derive dynamic output feedback controllers that render the closed-loop system externally positive. First, we introduce an equivalent representation for discrete-time, linear, time-invariant systems in the behavioral space, where we leverage the approach in \cite{AAAM-VK-VK-FP:22} to express any dynamic controller as a static behavioral feedback gain. After that, we use the behavioral representation of the system along with the results in \cite{EG-LN:15} to derive a behavioral feedback controller that monotonically tracks a constant non-negative reference output. Moreover, we derive a direct data-driven approach to obtain a behavioral feedback controller that monotonically tracks a constant non-negative reference output for SISO systems using input-output data. Both approaches, model-based and data-based, allow us to obtain output-feedback controllers that render the closed-loop system externally positive. Finally, we validate our results numerically in a drone landing control problem. Our approach is based on arbitrarily placing the closed-loop eigenvalues such that the closed-loop system is stable and externally positive. Several future directions can be explored, including optimal placement of the eigenvalues to optimize a specified performance metric, as well as, improving the robustness of the derived controllers against perturbations, such that the closed-loop system remains stable and externally positive when deployed in noisy environment.
\appendix
\setcounter{section}{7} 
%
\subsection{System representation in the behavioral space}\label{app: behavioral dynamics}
The following Lemma provides an equivalent representation of \eqref{eq: system in x} in the behavioral space, $z$, which is written in~\eqref{eq: z dynamics}.

\begin{appxlem}{\bf \emph{(Equivalent dynamics \cite[Lemma 5.2]{AAAM-VK-VK-FP:22})}}\label{lemma: system in z}
  Let $z$ be as in~\eqref{eq: z}. Then, the evolution of $z$ is written as \eqref{eq: z dynamics}, where $\left.\mc{A}_u \triangleq\mc{F}_2-CA^{n}\mc{O}^{\dagger} \mc{F}_1\right.$, and $\left. \mc{A}_y\triangleq CA^{n}\mc{O}^{\dagger}\right.$, with
  
  %
  \begin{align*}
    \mc{O}&\triangleq \left[\begin{smallarray}{c}
      C\\
      CA\\
      \vdots \\
      CA^{n-1}
    \end{smallarray}\right], \quad 
    \mc{F}_1\triangleq\left[\begin{smallarray}{cccc}
      0    & \cdots & 0 & 0\\
      CB & \cdots & 0 & 0\\
      \vdots  & \ddots & \vdots & \vdots\\
      CA^{n-2}B & \cdots & CB & 0
    \end{smallarray}\right],\\
    \mc{F}_2&\triangleq\left[\begin{smallarray}{cccc}
      CA^{n-1} B & \cdots & CB
    \end{smallarray}\right].
  \end{align*}~\oprocend
%
\end{appxlem}
%
Note that system's observability is required in Lemma~\ref{lemma: system in z}. The proof of Lemma~\ref{lemma: system in z} follows similar steps as that of \cite[Lemma 5.2]{AAAM-VK-VK-FP:22}.
\subsection{Verifying the condition in Theorem \ref{thrm: equivalence}}\label{app: equivalence}
In this Appendix, we verify the condition in Theorem \ref{thrm: equivalence}. In particular, we provide a method to compute an input sequence $\left. u_{0:n-1}\!\triangleq [u(0)^{\T},\!\cdots\!,\! u(n-1)^{\T}]^{\T} \right.$ such that the corresponding output sequence $y_{d:n+d-1}\!\triangleq \![y(d)^{\T},\!\cdots\!, y(n+d-1)^{\T}]^{\T}$ is non-negative for both the model-based and the data-based~settings.
%
%
\subsubsection{Model-based}
Given system \eqref{eq: system in x} with relative degree $d$. The~sequence $y_{d:n+d-1}$ can be written in the following form
\begin{align}\label{eq: yd sequence}
y_{d:n+d-1}=\mc{O}_d x(0) + \mc{F}_d u_{0:n-1},
\end{align}
where
\begin{align*}
\mc{O}_d&\triangleq \left[\begin{smallarray}{c}
      CA^d\\
      \vdots \\
      CA^{d+n-1}
    \end{smallarray}\right], \quad 
    \mc{F}_d\triangleq\left[\begin{smallarray}{ccccc}
      CA^{d-1} B   & \cdots & 0\\
      \vdots  & \ddots & \vdots  \\
      CA^{d+n-2}B & \cdots & CA^{d-1} B
    \end{smallarray}\right].
\end{align*}
We can choose $u_{0:n-1}$ such that $y_{d:n+d-1}=v\geq 0$ as:
\begin{align}\label{eq: desired input}
 u_{0:n-1}=\mc{F}_d^{\dagger}\left(v-\mc{O}_dx(0)\right)+w,
\end{align}
where $w \in \Ker{\left(\mc{F}_d\right)}$. Note that the matrix $\mc{F}_d$ is full-row rank since the system is right invertible (Assumption \ref{assump: reference tracking}). Also, the vector $w$ can be arbitrarily chosen since upon plugging $u_{0:n-1}$ in \eqref{eq: desired input} into \eqref{eq: yd sequence}, the vector $w$ disappears ($\mc{F}_d w=0$). Therefore, we can choose~$w=0$.
\smallskip
\subsubsection{Data-based}
Consider $N \geq 2n$ input-output trajectories of length $n$, which are collected by applying persistently exciting input~to SISO system \eqref{eq: system in x} that satisfies Assumption \ref{assump: reference tracking} with relative degree $d$, and starting from arbitrary initial state. The data is written~as
\begin{align}\label{eq: data n}
\begin{split}
 U_N&\triangleq 
\left[\begin{smallarray}{ccc}
u^{1}_{0:n-1} & \cdots & u^{N}_{0:n-1}
    \end{smallarray}\right], 
 \quad 
X_0 \triangleq 
\left[\begin{smallarray}{ccc}
x^{1}(0) & \cdots & x^{N}(0)
    \end{smallarray}\right], \\
    Y_N & \triangleq 
\left[\begin{smallarray}{ccc}
y^{1}_{d:n+d-1} & \cdots & y^{N}_{d:n+d-1}
    \end{smallarray}\right],
    \end{split}
\end{align}
where $u^{i}_{0:n-1}$, $x^{i}(0)$, and $y^{i}_{d:n+d-1}$ denote the input, the initial state, and the corresponding output of the $i$-th trajectory for $i=\{1,\cdots , N\}$.
Since $u$ is persistently exciting, $U_N$ is full-row rank and we have $\Rank{\left[\begin{smallarray}{c}U_N\\X_0\end{smallarray}\right]}=2n$, then, using \cite[Lemma 2]{CDP-PT:19}, we can write any input-output trajectory with a specified initial state as a linear combination of the columns of $U_N$, $Y_N$, and $X_0$, respectively. In particular, for any input sequence $u_{0:n-1}$, initial state $x_0$, and the corresponding output $y_{d:n+d-1}$, we can write
\begin{align}\label{eq: linear comb}
 u_{0:n-1}=U_N \alpha, \quad y_{d:n+d-1}=Y_N \alpha, \quad x_0=X_0 \alpha,
\end{align}
where $\alpha \in \mathbb{R}^{N}$. Our objective is that given an initial state $x_0$, we want to find a sequence $u_{0:n-1}$, such that $y_{d:n+d-1}\!\geq\! 0$, which boils down to choosing $\alpha$ such that $y_{d:n+d-1}\geq 0$. From \eqref{eq: linear comb}, we can write $\alpha\!=\! X_0^{\dagger}x_0+ X_{\text{null}}\mu$, where $X_{\text{null}}$ is a basis of $\Ker{\left(X_0\right)}$ and $\mu \in \mathbb{R}^n$ is an arbitrary vector, which can be~chosen~such~that $y_{d:n+d-1}\!\geq\!0$. In what follows, we make use of the following~Lemma.
\begin{appxlem}{\bf \emph{(Rank of $Y_NX_\Null$)}}\label{lemma: rank of YX_null}
Given the data in \eqref{eq: data n} with $X_\Null$ denoting the basis of $\Ker{\left(X_0\right)}$. Then
\begin{align*}
\Rank\left(Y_N X_\Null \right)=n.
\end{align*}
\begin{IEEEproof}
From \eqref{eq: yd sequence} we have
\begin{align*}
Y_N X_\Null=\mathcal{O}_dX_0 X_\Null +\mathcal{F}_d U_N X_\Null= \mathcal{F}_d U_N X_\Null.
\end{align*}
We have $\Rank\left(U_N \right)=n$ and the rows of $\left[\begin{smallarray}{c}U_N\\X_0\end{smallarray}\right]$ are linearly independent since the input is persistently exciting, and $\Rank \left(X_\Null \right)=N-n$ since $X_0$ is full-row rank{\footnote{This condition is typically satisfied for random choices of the initial states.}}. Hence, $\Rank\left(U_N X_\Null \right)=n$. Further, notice that for SISO system \eqref{eq: system in x} with relative degree $d$ that satisfies Assumption \ref{assump: reference tracking}, the matrix $\mc{F}_d$ is square and full-rank with $\Rank{\left(\mc{F}_d \right)}=n$. Therefore, we have $\Rank{\left(\mc{F}_d U_N X_\Null \right)}=n$.
\end{IEEEproof}
\end{appxlem}
\smallskip
Let $y_{d:n+d-1}=v\geq 0$, then from \eqref{eq: linear comb} we can write
\begin{align*}
 y_{d:n+d-1}&=Y_N \alpha=Y_N X_0^{\dagger}x_0+ Y_NX_{\text{null}}\mu=v,\\
 \implies \mu&=\left(Y_NX_{\text{null}}\right)^{\dagger}\left(v-Y_N X_0^{\dagger} x_0\right) +w,
\end{align*}
where $w \in \Ker{\left(Y_NX_{\text{null}}\right)}$. Then we can write $\alpha$ as 
\begin{align}\label{eq: alpha}
\alpha= X_0^{\dagger}x_0+ X_{\text{null}}\left(Y_NX_{\text{null}}\right)^{\dagger}\left(v-Y_N X_0^{\dagger} x_0\right)+X_{\text{null}}w,
\end{align}
Notice that from \eqref{eq: alpha}, \eqref{eq: linear comb}, and Lemma \ref{lemma: rank of YX_null}, for any choice of $w \in \Ker{\left(Y_NX_{\text{null}}\right)}$ we have $y_{d:n+d-1}=Y_N\alpha=v$, hence we can choose $w=0$. Finally, from \eqref{eq: alpha}, the input sequence $u_{0:n-1}\!=\!U_N\alpha$ that ensures $y_{d:n+d-1}\!=\!v\geq\!0$ can be written~as
\begin{align*}
 u_{0:n-1}=U_N X_0^{\dagger} x_0 +U_N X_{\text{null}} \left(Y_NX_{\text{null}}\right)^{\dagger}(v-Y_N X_0^{\dagger} x_0).
\end{align*}
%
%
%
%
%
%

\bibliographystyle{unsrt}
\bibliography{alias,Main,FP,New}

\end{document}